\documentclass[10pt,english,pre,nofootinbib,reprint,showpacs,showkeys,preprintnumbers,floatfix, longbibliography]{revtex4-1}
\usepackage[latin9]{inputenc}
\usepackage[a4paper]{geometry}
\usepackage{xcolor}
\usepackage{pdfcolmk}
\usepackage{graphicx}
\usepackage{babel}
\usepackage{mathrsfs}
\usepackage{amsmath}
\usepackage{amssymb}
\usepackage{esint}
\usepackage{subcaption}
\PassOptionsToPackage{normalem}{ulem}
\usepackage{ulem}
\usepackage{dsfont} 
\usepackage[unicode=true,pdfusetitle,
 bookmarks=true,bookmarksnumbered=false,bookmarksopen=false,
 breaklinks=false,pdfborder={0 0 1},backref=false,colorlinks=false]
 {hyperref}

\makeatletter

 \@ifundefined{textcolor}{}
 {%
   \definecolor{BLACK}{gray}{0}
   \definecolor{WHITE}{gray}{1}
   \definecolor{RED}{rgb}{1,0,0}
   \definecolor{GREEN}{rgb}{0,1,0}
   \definecolor{BLUE}{rgb}{0,0,1}
   \definecolor{CYAN}{cmyk}{1,0,0,0}
   \definecolor{MAGENTA}{cmyk}{0,1,0,0}
   \definecolor{YELLOW}{cmyk}{0,0,1,0}
 }

\usepackage{aas_macros}
\usepackage{bbold}

\let\textquotedbl="

\makeatother

\begin{document}

\title{Towards information optimal simulation of partial differential equations}

\author{Reimar H. Leike, Torsten A. Enßlin}

\affiliation{{\small{}Max-Planck-Institut für Astrophysik, Karl-Schwarzschildstr.~1,
85748 Garching, Germany}\\
Ludwig-Maximilians-Universität München, Geschwister-Scholl-Platz{\small{}~}1,
80539 Munich, Germany}
\begin{abstract}
Most simulation schemes for partial differential equations (PDEs) focus on minimizing a simple error norm of a discretized version of a field. This
paper takes a fundamentally different approach; the discretized field is interpreted as data providing information about a real physical field that is unknown. This information
is sought to be conserved by the scheme as the field evolves in time. Such an information theoretic approach to simulation was pursued before by
information field dynamics (IFD). In this paper
we work out the theory of IFD for nonlinear PDEs in a noiseless Gaussian approximation.
The result is an action that can be minimized to obtain an informationally optimal simulation scheme.
It can be brought into a closed form using field operators to calculate the appearing Gaussian integrals.
The resulting simulation schemes are tested numerically in two instances for the Burgers equation. Their accuracy surpasses finite-difference schemes
on the same resolution. The IFD scheme, however, has to be correctly informed on the subgrid correlation structure. In certain limiting cases we recover well-known
simulation schemes like spectral Fourier Galerkin methods. We discuss implications of the approximations made.
\end{abstract}

\keywords{information theory, Bayesian inference, simulation, partial differential equation}
\maketitle

\section{Introduction}

Simulation of partial differential equations (PDEs) is a wide field with countless applications. This stems from the fact that there is
no general known analytic solution for most of the interesting, in practice occurring PDEs. Thus one has to resort to simulation in order to make
predictions about the behavior of the solutions. PDEs are differential equations for fields, which have infinite degrees of freedom. However, on a computer one
is not able to store the whole field for any point in time. Furthermore the time evolution has to be discretized as well, because time is a 
continuous variable. 

Information field dynamics (IFD) \citep{2013PhRvE..87a3308E} takes an approach that differs slightly on a fundamental level from conventional field discretization. 
Instead of simulating a discretized field,
finite information about the real continuous field is stored in a data vector, as if it were obtained from a measurement. The time evolution of the data is then derived from the evolution of the real field.
This interpretation enables the application of information theory, specifically information field
theory \citep{1999physics..12005L, 2009PhRvD..80j5005E} which is information theory for the reconstruction of fields.

The application of information theory to get an improvement or better understanding of existing numerical methods is not new. One of the early prominent examples is Ref.\,\citep{diaconis1988bayesian},
where Bayesian inference is used to compute integrals. There are 
also examples of groups working on applying information theory to simulations. Some historic
examples are \citep{kailath1969general, joseph1968filtering,jazwinski1970mathematics, doucet1998sequential} who regard the problem as a hidden Markov model which is then treated in a Bayesian fashion
through a filtering approach. See e.g. \citep{bruno2013sequential} for an overview, \citep{ridgeway2003sequential} for a more generic review of sequential Monte Carlo methods. 
There is still ongoing research for the filtering approach, see e.g. \citep{branicki2014quantifying,sullivanbayesian, kersting2016active}. These
methods can also be applied to neural networks, see e.g. \citep{de2001sequential}.
In some cases, for example for linear differential equations, one can 
infer the solution directly \citep{raissi2017inferring}. Other approaches focus on parametrizing the posterior as a Gaussian and learning the dynamics
in a way motivated by machine learning \citep{raissi2017numerical}. 

The approach that is probably the closest to the one in this paper is described in \citep{archambeau2007gaussian}, where stochastic differential equations
are approximately solved using a variational approach. The differences to our approach lie in the way the probability density is
parametrized and how the KL divergence is used.

All in all, Bayesian simulation is an active and growing field of research.

In our approach, we do not rely on sampling, neither are we restricted to linear PDEs. Instead, we approximate the true evolved probability distribution
of solutions by a parametrized one in each time step; and choose the parameters so that the loss of information is minimized. Hereby we parameterize
the probability distribution such that it mimics a physical measurement instrument.

In Sec.\,\ref{sec:general-formalism} the reader is introduced to the fundamental concepts of IFD and general formula for discretizing PDEs.
This formula is then tested in Sec.\,\ref{sec:numerical-tests} for its numerical performance. 
Advantages and disadvantages of the proposed scheme are discussed in Sec.\,\ref{sec:current-caveats}. 
We conclude in Sec.\,\ref{sec:conclusion}.

\section{General formalism}
\label{sec:general-formalism}

IFD is a formalism for simulating differential equations for fields $s = s(x,t)$ of the form
\begin{align}
\frac{\text{d}s}{\text{d}t} = f(s) \label{eq:PDE}
\end{align}
using only the finite resources that are available on computers. 
This implies that from the infinite degrees of freedom of a field $s$, 
only finitely many can be taken into account. IFD is a specific kind of Bayesian forward simulation scheme. 

A forward simulation scheme takes
a data vector $d_0$ and returns a new data vector $d_1 = d_1(d_0)$. Here a data vector is referring
to an array of numbers on a computer. The data $d_0$ are supposed to contain information about the
real physical field $s_0$ at time $t_0$. What kind of information $d$ contains is also specified by the simulation scheme.
The new data vector $d_1$ is supposed to contain information about the field $s_1$ at the time $t_1 = t_0+\text{d}t$. One then iterates
the application of this scheme until one arrives at a target time. On this abstract level IFD yields a forward simulation scheme. The
difference to the construction of most other simulation schemes is its information theoretic foundation and very restrictive formalism.
The formalism is restrictive in the sense that once it is defined what information the data $d$ contains about the field $s$, the time
evolution of the scheme $d_1(d_0)$ is completely specified.

IFD attempts to mimic the optimal Bayesian simulation. In an optimal Bayesian simulation we take the knowledge
about the initial conditions $P(s_0|\text{``init'})$ and then compute the time evolved probability density using
the exact analytic time evolution
\begin{align}
 P(s_n|\text{``init''}) = P(s_0(s_n)|\text{``init''})\left|\left|\frac{\partial s_0(s_n)}{\partial s_n}\right|\right| \ .
\end{align}
Here we have assumed that there exists an exact solution for the PDE to be simulated (at least up to a zero set of $P(s_0|\text{``init''})$ ). Thus
there is a one to one mapping between fields $s_0$ at time $t_0$ and fields $s_n$ at time $t_n$, such that we can write the initial field $s_0(s_n)$ as function of the later field $s_n$,
or vice versa. We denote with $\left|\left|\frac{\partial s_0(s_n)}{\partial s_n}\right|\right|$ the absolute value of the Jacobi determinant that arises from transforming the probability density. Note that
no information is lost, since time forward and backward evolution is a one to one mapping of the phase space of the field $s$. This optimal Bayesian simulation scheme is
practically not accessible in most interesting cases because it requires the exact (backward) time evolution to be known, the Jacobi determinant to be computable and the storage of
whole probability densities over fields. In this paper, we propose a scheme that aims to overcome these limitations at the cost of losing some information in the process of simulation.
It does so by parameterizing the probability density
and then evolving these parameters such that the least amount of information is lost in each of the small time steps. We proceed by describing in detail how the probability density is
parametrized, then we describe how we approximate the time evolution.

In IFD we store a finite amount of data $d$ on the 
actual continuous field $s$, \emph{as if} measured by an instrument whose action is described by a measurement equation of the following form:
\begin{align}
d = R(s)(t)+n
\end{align}
Here $n$ is accumulated numerical noise and $R$ is some response function.
As an example for $R$ one could choose a matrix
of Dirac $\delta$-distributions for $R$ to mimic point measurements of the field at certain locations.
This measurement equation does not imply that there is an actual measurement, it just defines the
probability theoretic connection between the data $d$ on our computer and the actual physical field $s$ that
we want to simulate. The initial conditions of the PDE will determine the first data $d_0$, future
data will then be determined by the scheme to mimic the time evolution of the field $s$.
To recover the full field $s_0$ from the data $d_0$ at time $t_0$ one can use Bayes theorem:
\begin{align}
P(s_0|d_0) = \frac{P(d_0|s_0)P(s_0)}{P(d_0)} \ .
\end{align}
For this, a prior $P(s_0)$ is necessary. It reflects the knowledge
about $s_0$ when no data $d_0$ are available. A simulation scheme also has to discretize
the time evolution such that in a time step from $t_0$ to $t_1=t_0+\text{d}t$ the data gets updated from $d_0$ to $d_1$. 
In the language just introduced, the purpose of
a simulation scheme is to choose a proper time discretization $d_1(d_0)$ that
is as close to the real evolution of $U(s_0)=s_1$ as possible (or even equal 
if feasible) and a proper discretization $R$ of space such that the
features of the field are represented well.

In IFD the time evolution of the data $d$ is defined indirectly, that is we assign $d_1$ such that the posterior $P(s_1|d_1)$ using our new data $d_1$
matches the time evolved probability density $P(s_1|d_0)$ using our old data $d_0$ as well as possible. Note that $P(s_1|d_0)$ is what we defined to be
the optimal Bayesian simulation, but simulated only for a small time step $\text{d}t$, where a linearization of the time evolution might still be justified. Because we cannot
store the whole density $P(s_1|d_0)$ we store an approximation of it that uses the same parametrization as the probability density $P(s_0|d_0)$ but with new values $d_1$ assigned
to the parameters such that it approximates the time evolved probability density $P(s_1|d_0)$ as well as possible. For probability densities corresponding to a Bayesian belief, there is only
one consistent notion of ``approximating as well as possible'', given the two requirements that the optimal approximation is no approximation and
that an approximation can be judged by what it predicts for actual outcomes. 
We refer to Ref.\,\citep{2016arXiv161009018L} for a practice-oriented discussion why this uniquely determines the ``approximation'' KL distance as the appropriate loss function to be used here,
see Ref.\,\citep{bernardo1979expected} for the original proof on probability densities.
This proposed loss is different than that in the originally proposed IFD scheme \citep{2013PhRvE..87a3308E} and leads to matching the two distributions via
\begin{align}
\text{KL}(d_0,d_1) = \int \text{d}s_1\, P(s_1|d_0)\text{ln}\frac{P(s_1|d_0)}{P(s_1|d_1)}\label{eq:futureKL}
\end{align}
In this matching, $d_0$ is given and $d_1$ is searched for, such that the KL divergence serves as an action that is minimized to obtain the discretized time
evolution $d_1(d_0)$.

It was also proposed in \citep{2016arXiv161009018L} that for information preserving dynamics, that is for non-stochastic time evolution, one has 
$P(s_1|d_0)=P(s_0|d_0)\left|\left|\frac{\partial s_0(s_1)}{\partial s_1}\right|\right|$
and therefore this Kullback-Leibler distance is
equal to the Kullback-Leibler distance with both probability densities time evolved to the past (note the changed indices):
\begin{align}
\text{KL}(d_0,d_1) &= \int \text{d}s_0\, P(s_0|d_0)\text{ln}\frac{P(s_0|d_0)}{P(s_0|d_1)}\label{eq:pastKL}
\end{align}
Note that the equality between Eqs.\,(\ref{eq:futureKL}) and (\ref{eq:pastKL})
is nothing else than the invariance of the KL under invertible transformations. In this case the transformation is the time evolution of the field $s$.
The latter KL can be calculated once one made a suitable choice for $R$.
For this note that there is a degeneracy between $R$ and $s$. That means, if $R$ is altered by an invertible operator $T$
\begin{align}
R^\prime = RT\label{eq:new-response}
\end{align}
then this is equivalent to instead simulating the differential equation for $Ts$
\begin{align}
\frac{\text{d}}{\text{d}t}(Ts) = \frac{\text{d}T}{\text{d}s}\left(f(T^{-1}(Ts))\right) \label{eq:transform-field}
\end{align}
and using the unaltered response $R$. This is because
\begin{align}
(R^\prime)s(t) = (RT)s(t) = R (Ts)(t)\ . \label{eq:gauge-freedom}
\end{align}
This provides some freedom to simplify $R$, thus $R$ can be chosen such that it is
linear at the cost of possibly making the time evolution more complicated.
If the prior $P(s_0)=\mathscr{G}(s_0,S_0)$ and the noise $P(n_0)=\mathscr{G}(n_0,N_0)$ are zero-centered 
Gaussian distributions with covariance matrices $S_0$ and $N_0$, respectively, then the inverse problem 
can be solved by a (generalized) Wiener Filter \citep{1949wiener} and a Gaussian posterior distribution is obtained:
\begin{align}
P(s_0|d_0) &= \mathscr{G}(s_0-m_0,D_0)\nonumber\\
&= \left|2\pi D_0\right|^{-\frac{1}{2}}e^{-\frac{1}{2}\left(s_0-m_0\right)^\dagger D_0^{-1}\left(s_0-m_0\right)}
\end{align}
Here $m_0 = D_0R^\dagger N_0^{-1} d_0$ and $D_0^{-1} = S_0^{-1}+R^\dagger N_0^{-1}R$.
One also gets a Gaussian posterior distribution for $s_1$:
\begin{align}
P(s_1|d_1) &= \mathscr{G}(s_1-m_1,D_1),\\
\text{with } D_1^{-1} &= S_1^{-1}+R^\dagger N_1^{-1}R\\
\text{and } m_1 &= D_1R^\dagger N_1^{-1}d_1 \label{eq:field-WF}\\
&= SR^\dagger \left(RS_1R^\dagger+N_1\right)^{-1}d_1\label{eq:data-WF}\ .
\end{align}
Note that Eqs.\,(\ref{eq:field-WF}) and (\ref{eq:data-WF}) are two equivalent ways to obtain a reconstruction $m_1$.
In our paper we will mostly use Eq.\,(\ref{eq:data-WF}) as the matrix inversion only needs to be computed for
$RS_1R^\dagger+N_1$, which is a finite dimensional operator.

To compute the necessary quantities for our action as given by Eq.\,(\ref{eq:pastKL}) we have to
compute the distribution for $s_0$ given $d_1$. It is obtained from the backward time evolution of $P(s_1|d_1)$: 
\begin{align}
P(s_0|d_1) = \mathscr{G}(U(s_0)-m_1,D_1)\left|\left|\frac{\text{d}U(s_0)}{\text{d}s_0}\right|\right|
\end{align}
Here $U(s(t_0)) = s(t_1)$ is the exact analytical time evolution.
Using this, the Kullback-Leibler divergence that needs to be minimized so that $d_1$ is obtained is
\begin{align}
\text{KL}(d_0,d_1)&=\int \text{d}s_0\, P(s_0|d_0)\text{ln}\frac{P(s_0|d_0)}{P(s_0|d_1)}\\
&= \int \text{d}s_0\, \mathscr{G}(s_0-m_0,D_0)\nonumber\\
& \quad\text{ln}\frac{\mathscr{G}(s_0-m_0,D_0)}{\mathscr{G}(U(s_0)-m_1,D_1)\left|\left|\frac{\text{d}U(s_0)}{\text{d}s_0}\right|\right|} \nonumber\ .
\end{align}
We only minimize for parameters of $\mathscr{G}(U(s_0)-m_1,D_1)$, so we can ignore any additive terms that do not depend on $d_1$.
Thus
\begin{align}
&\text{KL}(d_0,d_1)\,\widehat{=}\nonumber\\ 
& \int \text{d}s_0\, \mathscr{G}(s_0-m_0,D_0)\,\text{ln}\frac{1}{\mathscr{G}(U(s_0)-m_1,D_1)} \ .
\end{align}
Here ``$\widehat{=}$'' denotes equality up to irrelevant constants, which in this case are constants that are not a function of $d_1$. These will
drop out when the expression is minimized with respect to $d_1$ later on.
Note that the absolute value of the Jacobian $\left|\left|\frac{\text{d}U(s_0)}{\text{d}s_0}\right|\right|$ can be ignored because it only depends on $s_0$.
The integral above can be quite difficult to evaluate in general. For integrals involving Gaussian distributions there is however a general method \citep{2016PhRvE..94e3306L}
to write down a closed expression for the result. Replacing every instance of $s_0$ with the field operator 
\begin{align}
O_{m_0} = m_0 + D_0\frac{\text{d}}{\text{d}m_0} 
\end{align}
allows us to evaluate the integral at the cost of having to evaluate operator expressions.
The integral is rewritten as
\begin{align}
\text{KL}(d_0,d_1) \,&\widehat{=}\,\text{ln}\frac{1}{\mathscr{G}(U(O_{m_0})-m_1,D_1)}\nonumber\\
&\widehat{=}\,\frac{1}{2}(U(O_{m_0})-m_1)^\dagger D_1^{-1}(U(O_{m_0})-m_1)\nonumber\\
&\ +\frac{1}{2}\text{tr}\left(\text{ln}\left(2\pi D_1\right)\right)\ . \label{eq:action}
\end{align}
We now minimize this Kullback-Leibler divergence with respect to $d_1$. For this we
compute the derivative
\begin{align}
\frac{\text{d}\text{KL}(d_0,d_1)}{\text{d}d_1} &= \left(\frac{\text{d}m_1}{\text{d}d_1}\right)^\dagger D_1^{-1}(m_1-U(O_{m_0}))\nonumber\\
&= N_1^{-1}R(m_1-U(O_{m_0}))\label{eq:gradient-m}
\end{align}
with respect to $d_1$. We now assume that we leave the noise matrix, the response, and the prior invariant, thus omitting the indices on
these operators.
At the minimum this derivative is $0$, so we can solve it for $d_1$:
\begin{align}
0&= N^{-1}R(m_1-U(O_{m_0}))\nonumber\\
0&= RSR^\dagger\left(N+RSR^\dagger\right)^{-1}d_1-RU(O_{m_0})\nonumber\\
d_1&= \left(N+RSR^\dagger\right)\left(RSR^\dagger\right)^{-1}RU(O_{m_0})\label{eq:d_1(d_0)}
\end{align}
One way to use IFD is to reformulate a PDE like Eq.\,(\ref{eq:PDE}) to an ordinary differential equation (ODE), for which potent solvers already exist. For this we expand
Eq.\,(\ref{eq:d_1(d_0)}) to first order in $\text{d}t$:
\begin{align}
d_1&= \left(N+RSR^\dagger\right)\left(RSR^\dagger\right)^{-1}R\left(O_{m_0}+\text{d}tf(O_{m_0})\right)\nonumber\\
d_1&= \left(N+RSR^\dagger\right)\left(RSR^\dagger\right)^{-1}R\nonumber\\
&\quad\left(SR^\dagger\left(N+RSR^\dagger\right)^{-1}d_0+\text{d}tf(O_{m_0})\right)\nonumber\\
d_1&= d_0+\left(N+RSR^\dagger\right)\left(RSR^\dagger\right)^{-1}R\text{d}tf(O_{m_0})
\end{align}
Inserting $d_1 = d_0 + \text{d}t\frac{\text{d}d}{\text{d}t}$ we arrive at an ODE for $d$:
\begin{align}
\frac{\text{d}d}{\text{d}t}&= \left(N+RSR^\dagger\right)\left(RSR^\dagger\right)^{-1}Rf(O_{m})\label{eq:central-with-noise}
\end{align}
Using this in the limit of no-noise $N\rightarrow 0$ we get the following compact expression for the updating rule:
\begin{align}
\frac{\text{d}d}{\text{d}t} &= Rf(O_{m})  = Rf(O_{SR^\dagger\left(R S R^\dagger\right)^{-1}d}) \label{eq:central}
\end{align}
Eqs.\,(\ref{eq:central-with-noise}) and (\ref{eq:central}) are the central equations of this paper,
allowing us to discretize any differential equation. They were derived through minimizing the action given by Eq\,(\ref{eq:pastKL})
and thus mimic the Bayes optimal simulation up to a minimized information loss.
Using Eqs.\,(\ref{eq:central-with-noise}) and (\ref{eq:central}) and an appropriate choice of the response $R$ of the virtual measurement connecting field and data,
IFD tells us how the differential operators need to be discretized.

\section{Numerical tests}
\label{sec:numerical-tests}

As a benchmark we simulate the Burgers equation
\begin{align}
 \frac{\partial s}{\partial t}  = f(s) =  \eta\frac{\partial^2 s}{\partial x^2} - s\frac{\partial s}{\partial x}\ . 
\end{align}
This equation is numerically challenging as it develops shock waves for small diffusion constants $\eta$.
First, we have to specify our choice of $R$. We demonstrate the formalism for two different choices of $R$.

\subsection{Box grid \label{sec:box-grid}}
\begin{figure*}[t]
  \begin{minipage}[t]{\columnwidth}
   \includegraphics[width=.95\textwidth]{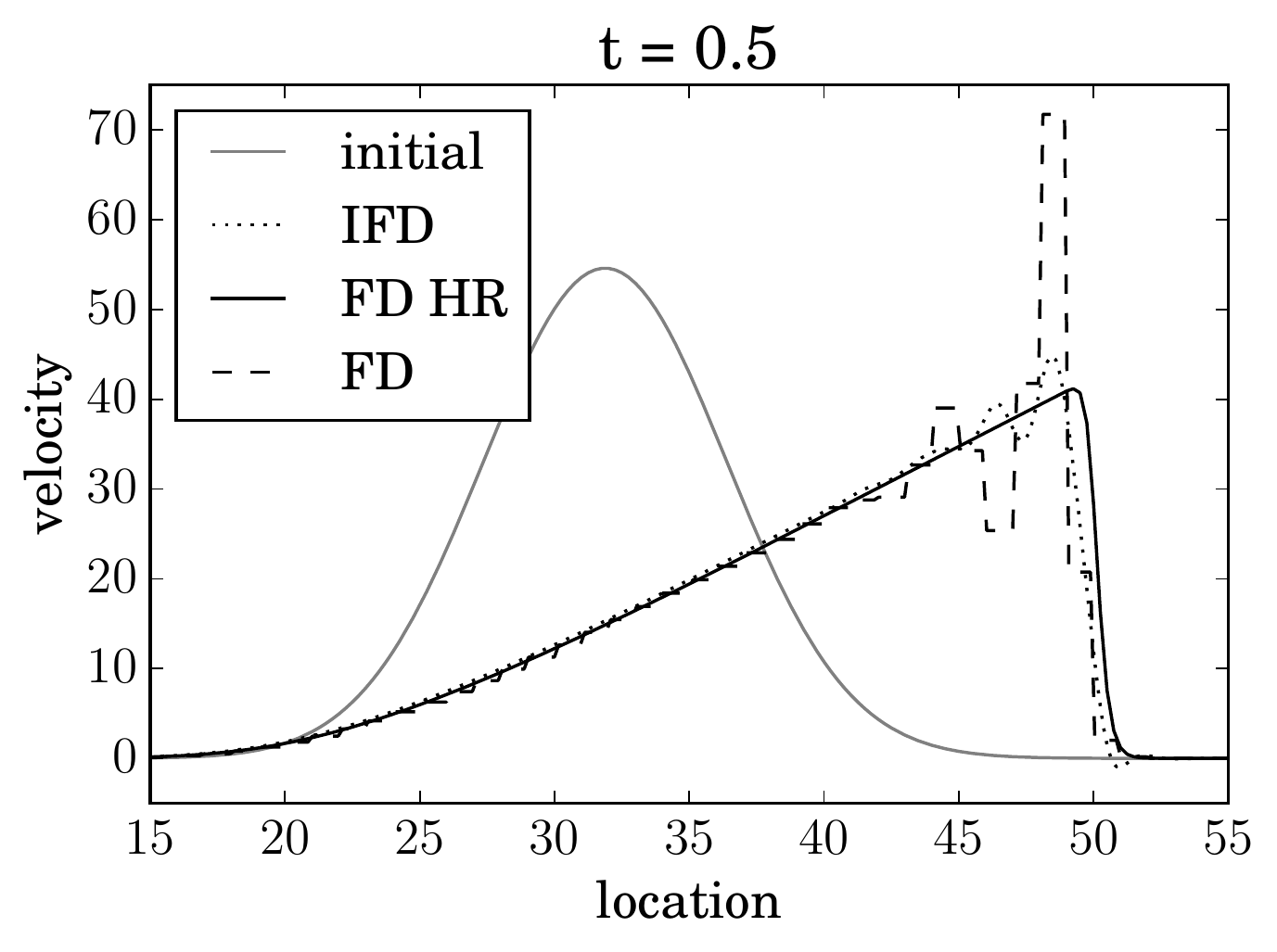}
    \caption{  
    Simulation of the Burgers equation using a Gaussian velocity profile as initial condition as represented by the gray line. The dotted line shows the reconstruction as it is obtained from the IFD formalism, the dashed line shows 
    a finite-difference simulation with the same resolution. The solid line is a more exact simulation obtained by simulating with a finite-difference scheme in 4-times higher spatial resolution.}
   \label{fig:box-sim}
  \end{minipage}
  \hfill
  \begin{minipage}[t]{\columnwidth}
    \includegraphics[width=.94\textwidth]{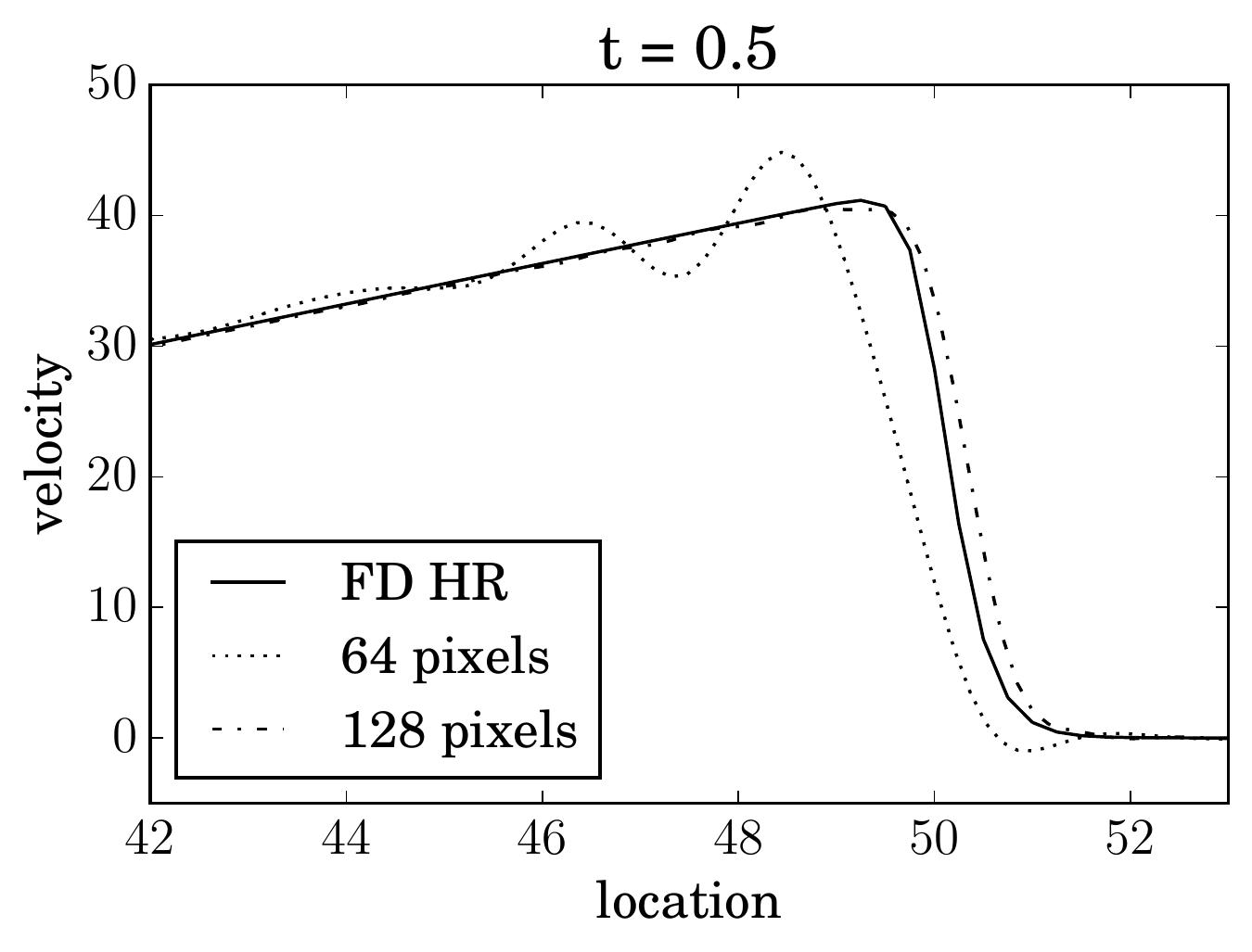}
    \caption{  
    Zoom in into the simulation of the Burgers equation shown in Fig.\,\ref{fig:box-sim}. The dash-dotted line shows the reconstruction as it is obtained from the IFD formalism,
    but with double the spatial resolution. The other lines are the same as in Fig.\,\ref{fig:box-sim}.}
   \label{fig:resolution-sim}
  \end{minipage}
\end{figure*}

We choose 
\begin{align}
R_{ix} = 1_{x_i,x_{i+1}}=
\begin{cases}
1 & x_i < x \leq x_{i+1}\\
0 & \text{otherwise.}
\end{cases}
\end{align}
This type of grid is commonly used in simulations. Starting from equation (\ref{eq:central}), we compute:
\begin{align}
\frac{\text{d}d}{\text{d}t} &= Rf(O_{m}) \nonumber\\
 &=R(\Delta O_{m} - O_m \nabla O_{m})\nonumber\\
 &=R(\Delta m - m\nabla m) \nonumber\\
 &\quad- \int\text{d}x\text{d}y\text{d}z\,R_{\cdot x}D_{xy}\frac{\text{d}}{\text{d}m_y}(\nabla_{xz} m_{z}) \nonumber\\
 &=R\Delta S R^\dagger (RSR^\dagger)^{-1}d\nonumber\\
 &\quad - R\left(S R^\dagger (RSR^\dagger)^{-1}d (\nabla S R^\dagger (RSR^\dagger)^{-1}d)\right)\nonumber\\
 &\quad - \int\text{d}x\text{d}y\,R_{\cdot x}D_{xy}\nabla_{xy} \label{eq:box_all_terms}
\end{align}
We introduce the short hand notation
\begin{align}
d^\prime = (RSR^\dagger)^{-1}d
\end{align}
so that the IFD Burgers scheme simplifies to
\begin{align}
 \frac{\text{d}d}{\text{d}t} &=R\Delta S R^\dagger d^\prime\nonumber\\
 &\quad - R\left(S R^\dagger d^\prime (\nabla S R^\dagger d^\prime)\right)\nonumber\\
 &\quad - \int\text{d}x\text{d}y\,R_{\cdot x}D_{xy}\nabla_{xy} \label{eq:box_all_terms_dprime}
\end{align}
Assuming that our a priori knowledge favors no certain points in space or certain directions, according to the Wiener-Khintchin theorem \citep{wiener1930generalized} 
the covariance operator $S$ has to be diagonal in Fourier space.
This is equivalent to a convolution with a convolution kernel $C_x$ in configuration ($x$-) space, such that
\begin{align}
 (SR^\dagger)_{xi} = C_x\ast R^\dagger_{xi} = \int\text{d}y\, C_{x-y} R^\dagger_{yi}\label{eq:defconv}\ .
\end{align}

We now compute the three terms of Eq.\,(\ref{eq:box_all_terms_dprime}) all separately, starting with the term involving the Laplace operator:
\begin{align}
&\left(R\Delta S R^\dagger d^\prime\right)_i = \int_{x_i}^{x_{i+1}} \text{d}x\, \sum_j\Delta \int_{x_j}^{x_{i+j}} \text{d}y\, S_{xy} d^\prime_{j}\nonumber\\
&= \int_{x_i}^{x_{i+1}} \text{d}x\, \sum_j\Delta C_{x} \ast 1_{x_j,x_{j+1}} d^\prime_{j}\nonumber\\
&= \int \text{d}x\,1_{x_i,x_{i+1}} \sum_j\Delta C_{x} \ast 1_{x_j,x_{j+1}} d^\prime_{j}\nonumber\\
&= \int_{x_i}^{x_{i+1}} \text{d}x\,  \sum_j\nabla \left(C_{x} \ast (\delta(x-x_j)-\delta(x-x_{j+1}))\right) d^\prime_{j}\nonumber\\
&= \int\text{d}x\,(\delta(x-x_{i+1})-\delta(x-x_i))\nonumber\\
&\qquad\sum_j\left(C_{x-x_j}-C_{x-x_{j+1}}\right) d^\prime_{j}\nonumber\\
&= \sum_j\left(C_{x_{i+1}-x_j}-C_{x_{i+1}-x_{j+1}}\right)\nonumber\\
&\qquad-\sum_j\left(C_{x_i-x_j}-C_{x_i-x_{j+1}}\right) d^\prime_{j}\nonumber\\
&= \sum_j\left(C_{l(i-j+1)}-2C_{l(i-j)}+C_{l(i-j-1)}\right) d^\prime_{j} \label{eq:box_int1}
\end{align}
Here we assumed the $x_i$ to be equally spaced with distance $l$. Note that this version of the discretized Laplace operator has similarities with the normal finite-difference \citep{Courant1928} Laplace operator, 
but accounts for the field correlation structure. We continue by computing the second term
\begin{align}
&\left(R\left(S R^\dagger d^\prime (\nabla S R^\dagger d^\prime)\right)\right)_i=\nonumber\\
&=\int\text{d}x\, R_{ix}(S R^\dagger d^\prime)_x(\nabla S R^\dagger d^\prime)_x\nonumber\\
&=-\int\text{d}x\,\left[\nabla\left( R_{ix}\left(S R^\dagger d^\prime\right)\right)_x\right](S R^\dagger d^\prime)_x\nonumber\\
&=-\int\text{d}x\,(\nabla R_{ix})\left(S R^\dagger d^\prime\right)(S R^\dagger d^\prime)\nonumber\\
&\qquad-\int\text{d}x\,R_{ix} \left(\nabla S R^\dagger d^\prime\right)(S R^\dagger d^\prime)\ .\label{eq:phoenix}
\end{align}
The last summand in Eq.\,(\ref{eq:phoenix}) is the same term we started with, only with a negative sign. Thus we can bring both to the same side of the equation and get
\begin{align}
&R\left(S R^\dagger d^\prime (\nabla S R^\dagger d^\prime)\right)=\nonumber\\
&=-\frac{1}{2}\int\text{d}x\, (\nabla R_{ix}) \left(S R^\dagger d^\prime\right)(S R^\dagger d^\prime)\nonumber\\
&= \frac{1}{2}\left[\left(C_x \ast\sum_j R_{xj}d^\prime_j\right)\left(C_x \ast \sum_k R_{xk}d^\prime_k\right)\right]_{x=x_i}^{x_{i+1}}\label{eq:box_int2}\ .
\end{align}
The third term is
\begin{align}
\int\text{d}x\text{d}y\,R_{ix}D_{xy}\nabla_{xy} \ .
\end{align}
This term vanishes in the case of periodic boundary conditions. One can see this by rewriting $\nabla_{xy} = \epsilon^{-1}\left(\delta(x-y+\epsilon)-\delta(x-y-\epsilon)\right)$
for a sufficiently small $\epsilon$ to obtain
\begin{align}
&\int\text{d}x\text{d}y\,R_{ix}D_{xy}\epsilon^{-1}\left(\delta(x-y+\epsilon)-\delta(x-y-\epsilon)\right)\nonumber\\
&=\int\text{d}x\,\epsilon^{-1}R_{ix}\left(D_{x(x+\epsilon)}-D_{x(x-\epsilon)}\right)\ .
\end{align}
Because $S$ and $R$ have no favored direction, $D_{x(x+\epsilon)} = D_{x(x-\epsilon)}$ and thus the third term vanishes.
Finally we have to compute 
\begin{align}
d^\prime &= \left(RSR^\dagger\right)^{-1} d\nonumber\\
&= \left(\int_{x_i}^{x_{i+1}}\int_{x_j}^{x_{j+1}}\text{d}x\text{d}y \, C_{x-y}\right)^{-1}d_j\label{eq:box_int3}\ .
\end{align}
Now that one has all the terms of Eq.\,(\ref{eq:box_all_terms}) one can choose a prior and obtain a simulation scheme as a result. Normally, one would choose the prior
according to physical properties of the system, such that it meaningfully encodes our knowledge in the absence of data. The matter of choosing priors will
be further addressed in Sec.\,\ref{sec:prior-choosing}. We just want to demonstrate the formalism, so we simply choose the analytic form of $C_x$ 
such that we can easily compute the three integrals given by 
Eqs.\,(\ref{eq:box_int1},\,\ref{eq:box_int2},\,\ref{eq:box_int3}). One convenient choice of $C_x$ is a Gaussian\footnote{or a mixture of Gaussians}, for which we know all the above terms analytically.
One might equally well choose any correlation function and do these integrations numerically. Because these integrals only have
to be done once, this does not significantly increase the computation time of the resulting simulation scheme.

Note that all computed operators are local, meaning that they fall off as $C_x$ falls off. Thus, they can be truncated at a certain distance and the whole simulation scales only linearly with the number of data points.

Fig.\,\ref{fig:box-sim} shows an example of a simulation that was performed using the scheme that was worked out in this section. As a comparison, the figure
also shows a simulation using the finite-difference method with the same spatial resolution. This simulation uses $64$ data values and the function
\begin{align}
s(x) = e^{4-(x/64-0.5)^2}
\end{align}
as initial condition. The simulation space is an interval of length $64$ with periodic boundary conditions. The prior covariance was chosen to be a convolution with a zero-centered Gaussian that has a standard deviation of $0.5$. The diffusion constant $\eta$ was chosen to be $5$. The result of a simulation using the finite-difference method
with a four times higher resolution is displayed as well. This high resolution simulation should capture all features produced by the Burgers dynamics. To investigate the resolution
dependence of the IFD scheme, we compare the IFD simulation of  Fig.\,\ref{fig:box-sim} with a simulation on a two times more resolved grid in Fig.\,\ref{fig:resolution-sim}. For
comparison one can again see the fine resolved finite-difference method. One can clearly see an increased performance as the resolution increases.

\subsection{Fourier grid}

\begin{figure*}[t]
 
    \includegraphics[width=.5\textwidth]{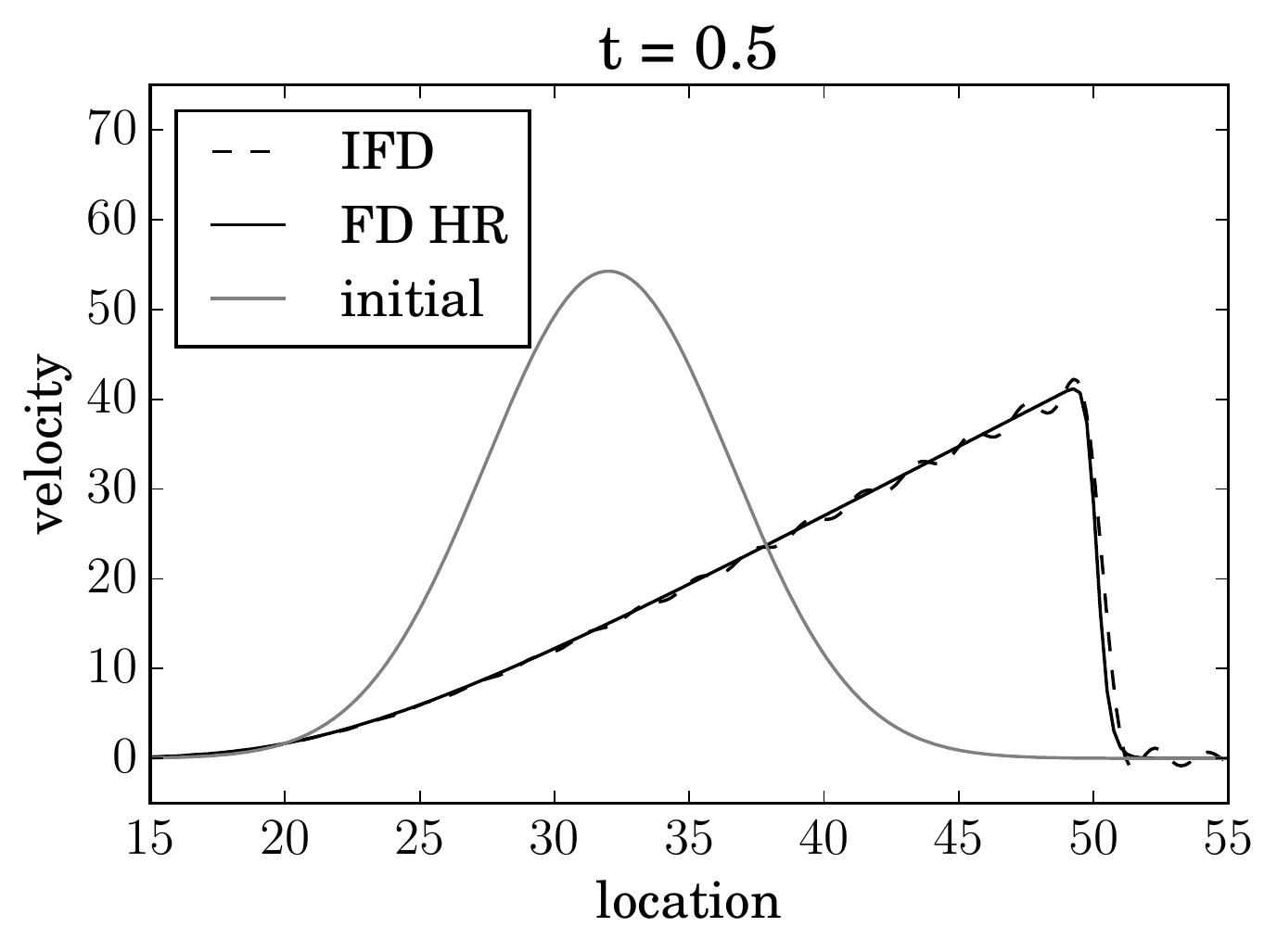}
    \caption{  
    Simulation of the Burgers equation using a Gaussian velocity profile as initial condition as represented by the grey line and a Fourier space response. 
    The dotted line is a reconstruction that is obtained through simulating with the IFD scheme. 
    This simulation scheme is equivalent to a Fourier-Galerkin scheme. The solid line is a high resolution simulation using a finite-difference scheme.}
   \label{fig:fourier-sim}
\end{figure*}

Now we switch to a different response.
We choose a Fourier space grid
\begin{align}
R_{ik} &= \sum_{i} \delta(k-k_i)
\end{align}
with Fourier space grid points $k_i$. There is a choice whether to view the Fourier transform
\begin{align}
F_{kx} = e^{ikx}
\end{align}
as part of the measurement $R$ or as transformation of the field $s_k = F_{kx}s_x$, see Eqs.\,(\ref{eq:new-response},\,\ref{eq:transform-field},\,\ref{eq:gauge-freedom}).
We choose the latter and obtain as transformed time evolution:
\begin{align}
\frac{\text{d}s_k}{\text{d}t} &= F_{kx}\frac{\text{d}s_x}{\text{d}t}\nonumber\\
&= F_{kx}\left(\Delta s_x-(s_x)(\nabla s_x)\right)\nonumber\\
&=  -k^2s_k+(s_k)\ast(ik s_k)
\end{align}
We insert this into Eq.\,(\ref{eq:central}) and obtain
\begin{align}
&\frac{\text{d}d}{\text{d}t} = Rf(O_{m})  = \nonumber\\
&-Rk^2SR^\dagger \left(RSR^\dagger\right)^{-1}d+ Rk^2SR^\dagger \left(RSR^\dagger\right)^{-1}d\ .
\end{align}
If the a priori knowledge does not favor any specific direction or point in space, then $S$ is diagonal in Fourier space. 
Thus 
\begin{align}
\left(SR^\dagger(RSR^\dagger)^{-1}\right)_{ki}&=\sum_{lrjp}S_{kl}\delta_{lm_j}(\delta_{m_ip} S_{pr}\delta_{rm_j})^{-1} \nonumber\\
&=\sum_{j}S_{km_j}(S_{m_im_j})^{-1} \nonumber\\
&= \delta_{km_i}=R^\dagger_{ki}
\end{align}
and the prior covariance drops out, making time evolution on this grid invariant under Fourier space priors.
For the final time evolution we arrive at
\begin{align}
\frac{\text{d}d_i}{\text{d}t} &= -k_i^2d_i\nonumber\\
&\ \,+\sum_j\int \text{d}k\text{d}k^\prime\,\delta(k-k_i-k^\prime)d_iik^\prime\delta(k^\prime-k_j)d_j\nonumber\\
&=-k_i^2d_i+\sum_j d_{j-i}ik_jd_j
\end{align}
where the continuous convolution was translated to a finite convolution on a grid. 
This resulting time evolution equation can be implemented efficiently and the results can be
seen in Fig.\,\ref{fig:fourier-sim}. The simulation constraints, initial conditions, and degrees of freedom were chosen to be the same as in Sec.\,\ref{sec:box-grid}.

The developed Fourier space IFD method is equivalent to a Fourier Galerkin method \citep{Galerkin1915}. In the Fourier
Galerkin method, the error to the correct solution is minimized for a vector in a subspace. This subset is a linear subspace of the real solution space, with selected basis functions that
are often chosen to be polynomials or, as in the case of the Fourier Galerkin method, Fourier basis functions. In the case of IFD this subspace is the co-image of the response $R$. 
Thus Galerkin schemes can be regarded as IFD schemes with $R$ being specified by the Galerkin basis. The prerequisites for this equivalence are that the prior $S$ commutes with the discretization $R^\dagger R$
and that we work in the no-noise regime $N\rightarrow 0$.

\section{Current advantages and disadvantages}
\label{sec:current-caveats}

In the current development status of the IFD method, there are some caveats as well as some advantages over classical approaches. 
Some of them stem from the theoretical side, where approximations had to be done in order to 
arrive at a computable scheme. In this section we discuss all the observed problems and benefits.

\subsection{Subgrid structure}

Information field dynamics employs the field evolution of the real physical field, and thus automatically takes into account subgrid structure. However, this leads to 
problems when the time evolution is truncated after the first order in $\text{d}t$. In most physical systems, the time evolution is faster for the smaller scales,
thus if we take into account all scales, no $\text{d}t$ is small enough to justify this approximation. In other schemes, the discretized field automatically yields a 
cutoff at high frequencies. However for IFD, in the derivation we truncate the whole time evolution of the real system at first order, which is not justified.
As an example consider the diffusion equation
\begin{align}
 \frac{\text{d}s}{\text{d}t} &= \Delta s \\
 \frac{\text{d}s_k}{\text{d}t} &= -k^2s_k\ .
\end{align}
For higher modes $k$ the change of $s$ scales with $k^2$. Thus, if we truncate the Taylor expansion of the time evolution at order of $\text{d}t$, then 
for $k>\sqrt{\frac{2}{\text{d}t}}$ we get
\begin{align}
 {s_k}^{t+\text{d}t} &= s_k^{t}-\text{d}t\,k^2s_k^t = -\left(1-\text{d}t\,k^2\right)s_k^{t}\ .
\end{align}
where $\left|1-\text{d}t\,k^2\right| > 1$, thus the scheme is boosting small frequencies instead of damping them.

This is, however, a general problem of any simulation scheme. 

\subsection{How to choose a prior\label{sec:prior-choosing}}

The goal of a prior is to incorporate as much information as one has about the system and nothing more. If the system has no special directions or singled out locations, 
one should choose a prior that is homogeneous and isotropic. These two
requirements force the covariance matrix $S_{xy} = \left<s_xs_y^\dagger\right>_{P(s)}$ to be diagonal in Fourier space. If we restrict ourselves to Gaussian priors,
the prior is fully characterized by its power spectrum
$P(k) \propto \left<\left|s_k\right|^2\right>_{P(s)}$. 
Thus, the
only a priori information that enters the simulation in the Gaussian case is how smooth the physical field is. But this is also a significant restriction. For example for infinitely sharp
shocks, as they occur in the Burgers equation with $\eta=0$, no smoothness at all is justified at the location of the shock, whereas at other points the solution might
be perfectly smooth. To capture this kind of behavior one would need to either use a prior that allows for higher order statistics, use
a dynamical prior which evolves in time, or to introduce data that capture the discontinuities.
In our simulation we observed that the scheme diverges quickly if an unjustified prior was chosen, that is for example a prior that enforces significantly more smoothness than 
is present in the solution.

To choose the right power spectrum of the prior one could use a fine grid simulation and take the occurring power spectrum as input for a coarser simulation.

\subsection{Static prior}

In the current formulation we assume that the prior does not evolve in time. However, ideally the prior should evolve with the system. This is
because some a priori assumptions that were made for time step $t_0$ might not be true at a later time (for example initial smoothness
might be violated by the formation of a shock). However, from an agnostic point of view, if one has no knowledge about points in time, then
the prior should be invariant under time translation.
In principle, IFD provides guidance on how to evolve any kind of degree of information on the field, like its covariance structure, and not only some measurement data. 
By minimizing Eq.\,\ref{eq:action} with respect
to any such piece of information, we obtain an evolution equation for it that loses the minimal amount of information. This way, when the prior is 
parametrized, an update rule for it is automatically obtained.

\subsection{No-noise approximation}

To derive our algorithms, we take the limit $N\rightarrow 0$. While this significantly simplifies the derivation of the schemes, it also deprives
the resulting schemes of the advantages that an information theoretic treatment has in general. In the no-noise approximation field
configurations $s$ with $Rs \neq d$ are assigned zero probability, thus the information loss in the presence of numerical rounding errors and
finite time steps, where the evolution of the data $d$ cannot satisfy Eq.\,(\ref{eq:central}) exactly, is infinite. 
Further development in the field of IFD will have to investigate into approaches incorperating noise.

\section{Conclusion}
\label{sec:conclusion}
The requirement of minimal information loss per time-step defines a unique simulation framework. 
This concrete simulation scheme requires the specification of the field measurement (response) and the incorporation of prior known correlation structure.
It exhibits similarities to the finite-difference scheme when the response is a grid of boxes
and becomes a spectral scheme in the case of Fourier space response. This yields a new interpretation for linear and in some cases even nonlinear Galerkin schemes.
These are information
optimal up to the approximations made in this paper if no spatial a priori knowledge about the field is available.

IFD can thus be regarded as a general theory for simulations that explains what assumptions about the simulated field enter a given simulation scheme, if one is 
able to reproduce that scheme in IFD. For some schemes one can enhance the performance by using a prior that is correctly
informed on the field correlation structure. When the prior is chosen incorrectly, for example if it is chosen such that the simulation produces features
that are regarded very unlikely by the prior, the scheme tends to diverge quickly.
In principle, IFD can provide a guideline how to evolve any degree of freedom; by minimizing
Eq.\,\ref{eq:pastKL} one gets a unique simulation scheme. An interesting route, at
least for the Burgers and other hydrodynamic equations, would be to automatically
infer the position of the virtual measurements, allowing the scheme to sample
the field where it is most informative. The investigation in that direction
is however beyond the scope of this paper and might be the target of future
research.

All in all information theory provides a powerful language to talk about simulation tasks. Even though the series of approximations made in this paper permitted
the resulting simulation schemes to only outperform finite differences by a small amount, further advancements in the field could yield substantial enhancements.



\section{Acknowledgements}
We would like to thank Martin Dupont, Phillip Frank, Sebastian Hutschenreuter, Jakob Knollm\"uller and two anonymous referees for the discussions and their valuable comments on the manuscript.

\bibliography{ift}

\begin{thebibliography}{24}%
\makeatletter
\providecommand \@ifxundefined [1]{%
 \@ifx{#1\undefined}
}%
\providecommand \@ifnum [1]{%
 \ifnum #1\expandafter \@firstoftwo
 \else \expandafter \@secondoftwo
 \fi
}%
\providecommand \@ifx [1]{%
 \ifx #1\expandafter \@firstoftwo
 \else \expandafter \@secondoftwo
 \fi
}%
\providecommand \natexlab [1]{#1}%
\providecommand \enquote  [1]{``#1''}%
\providecommand \bibnamefont  [1]{#1}%
\providecommand \bibfnamefont [1]{#1}%
\providecommand \citenamefont [1]{#1}%
\providecommand \href@noop [0]{\@secondoftwo}%
\providecommand \href [0]{\begingroup \@sanitize@url \@href}%
\providecommand \@href[1]{\@@startlink{#1}\@@href}%
\providecommand \@@href[1]{\endgroup#1\@@endlink}%
\providecommand \@sanitize@url [0]{\catcode `\\12\catcode `\$12\catcode
  `\&12\catcode `\#12\catcode `\^12\catcode `\_12\catcode `\%12\relax}%
\providecommand \@@startlink[1]{}%
\providecommand \@@endlink[0]{}%
\providecommand \url  [0]{\begingroup\@sanitize@url \@url }%
\providecommand \@url [1]{\endgroup\@href {#1}{\urlprefix }}%
\providecommand \urlprefix  [0]{URL }%
\providecommand \Eprint [0]{\href }%
\providecommand \doibase [0]{http://dx.doi.org/}%
\providecommand \selectlanguage [0]{\@gobble}%
\providecommand \bibinfo  [0]{\@secondoftwo}%
\providecommand \bibfield  [0]{\@secondoftwo}%
\providecommand \translation [1]{[#1]}%
\providecommand \BibitemOpen [0]{}%
\providecommand \bibitemStop [0]{}%
\providecommand \bibitemNoStop [0]{.\EOS\space}%
\providecommand \EOS [0]{\spacefactor3000\relax}%
\providecommand \BibitemShut  [1]{\csname bibitem#1\endcsname}%
\let\auto@bib@innerbib\@empty
\bibitem [{\citenamefont {{En{\ss}lin}}(2013)}]{2013PhRvE..87a3308E}%
  \BibitemOpen
  \bibfield  {author} {\bibinfo {author} {\bibfnamefont {T.~A.}\ \bibnamefont
  {{En{\ss}lin}}},\ }\bibfield  {title} {\enquote {\bibinfo {title}
  {{Information field dynamics for simulation scheme construction}},}\ }\href
  {\doibase 10.1103/PhysRevE.87.013308} {\bibfield  {journal} {\bibinfo
  {journal} {\pre}\ }\textbf {\bibinfo {volume} {87}},\ \bibinfo {eid} {013308}
  (\bibinfo {year} {2013})},\ \Eprint {http://arxiv.org/abs/1206.4229}
  {arXiv:1206.4229 [physics.comp-ph]} \BibitemShut {NoStop}%
\bibitem [{\citenamefont {{Lemm}}(1999)}]{1999physics..12005L}%
  \BibitemOpen
  \bibfield  {author} {\bibinfo {author} {\bibfnamefont {J.~C.}\ \bibnamefont
  {{Lemm}}},\ }\bibfield  {title} {\enquote {\bibinfo {title} {{Bayesian Field
  Theory: Nonparametric Approaches to Density Estimation, Regression,
  Classification, and Inverse Quantum Problems}},}\ }\href@noop {} {\bibfield
  {journal} {\bibinfo  {journal} {ArXiv Physics e-prints}\ } (\bibinfo {year}
  {1999})},\ \Eprint {http://arxiv.org/abs/physics/9912005} {physics/9912005}
  \BibitemShut {NoStop}%
\bibitem [{\citenamefont {{En{\ss}lin}}\ \emph {et~al.}(2009)\citenamefont
  {{En{\ss}lin}}, \citenamefont {{Frommert}},\ and\ \citenamefont
  {{Kitaura}}}]{2009PhRvD..80j5005E}%
  \BibitemOpen
  \bibfield  {author} {\bibinfo {author} {\bibfnamefont {T.~A.}\ \bibnamefont
  {{En{\ss}lin}}}, \bibinfo {author} {\bibfnamefont {M.}~\bibnamefont
  {{Frommert}}}, \ and\ \bibinfo {author} {\bibfnamefont {F.~S.}\ \bibnamefont
  {{Kitaura}}},\ }\bibfield  {title} {\enquote {\bibinfo {title} {{Information
  field theory for cosmological perturbation reconstruction and nonlinear
  signal analysis}},}\ }\href {\doibase 10.1103/PhysRevD.80.105005} {\bibfield
  {journal} {\bibinfo  {journal} {\prd}\ }\textbf {\bibinfo {volume} {80}},\
  \bibinfo {pages} {105005} (\bibinfo {year} {2009})},\ \Eprint
  {http://arxiv.org/abs/0806.3474} {arXiv:0806.3474} \BibitemShut {NoStop}%
\bibitem [{\citenamefont {Diaconis}(1988)}]{diaconis1988bayesian}%
  \BibitemOpen
  \bibfield  {author} {\bibinfo {author} {\bibfnamefont {Persi}\ \bibnamefont
  {Diaconis}},\ }\bibfield  {title} {\enquote {\bibinfo {title} {Bayesian
  numerical analysis},}\ }\href@noop {} {\bibfield  {journal} {\bibinfo
  {journal} {Statistical decision theory and related topics IV}\ }\textbf
  {\bibinfo {volume} {1}},\ \bibinfo {pages} {163--175} (\bibinfo {year}
  {1988})}\BibitemShut {NoStop}%
\bibitem [{\citenamefont {Kailath}(1969)}]{kailath1969general}%
  \BibitemOpen
  \bibfield  {author} {\bibinfo {author} {\bibfnamefont {Thomas}\ \bibnamefont
  {Kailath}},\ }\bibfield  {title} {\enquote {\bibinfo {title} {A general
  likelihood-ratio formula for random signals in gaussian noise},}\ }\href@noop
  {} {\bibfield  {journal} {\bibinfo  {journal} {IEEE Transactions on
  Information Theory}\ }\textbf {\bibinfo {volume} {15}},\ \bibinfo {pages}
  {350--361} (\bibinfo {year} {1969})}\BibitemShut {NoStop}%
\bibitem [{\citenamefont {Joseph}(1968)}]{joseph1968filtering}%
  \BibitemOpen
  \bibfield  {author} {\bibinfo {author} {\bibfnamefont {Peter~D}\ \bibnamefont
  {Joseph}},\ }\href@noop {} {\emph {\bibinfo {title} {Filtering for stochastic
  processes with applications to guidance}}}\ (\bibinfo {year}
  {1968})\BibitemShut {NoStop}%
\bibitem [{\citenamefont {Jazwinski}(1970)}]{jazwinski1970mathematics}%
  \BibitemOpen
  \bibfield  {author} {\bibinfo {author} {\bibfnamefont {Andrew~H}\
  \bibnamefont {Jazwinski}},\ }\bibfield  {title} {\enquote {\bibinfo {title}
  {Mathematics in science and engineering},}\ }\href@noop {} {\bibfield
  {journal} {\bibinfo  {journal} {Stochastic Processes and Filtering Theory}\
  }\textbf {\bibinfo {volume} {64}} (\bibinfo {year} {1970})}\BibitemShut
  {NoStop}%
\bibitem [{\citenamefont {Doucet}(1998)}]{doucet1998sequential}%
  \BibitemOpen
  \bibfield  {author} {\bibinfo {author} {\bibfnamefont {Arnaud}\ \bibnamefont
  {Doucet}},\ }\bibfield  {title} {\enquote {\bibinfo {title} {On sequential
  simulation-based methods for bayesian filtering},}\ }\href@noop {} {\
  (\bibinfo {year} {1998})}\BibitemShut {NoStop}%
\bibitem [{\citenamefont {Bruno}(2013)}]{bruno2013sequential}%
  \BibitemOpen
  \bibfield  {author} {\bibinfo {author} {\bibfnamefont {Marcelo~GS}\
  \bibnamefont {Bruno}},\ }\bibfield  {title} {\enquote {\bibinfo {title}
  {Sequential monte carlo methods for nonlinear discrete-time filtering},}\
  }\href@noop {} {\bibfield  {journal} {\bibinfo  {journal} {Synthesis Lectures
  on Signal Processing}\ }\textbf {\bibinfo {volume} {6}},\ \bibinfo {pages}
  {1--99} (\bibinfo {year} {2013})}\BibitemShut {NoStop}%
\bibitem [{\citenamefont {Ridgeway}\ and\ \citenamefont
  {Madigan}(2003)}]{ridgeway2003sequential}%
  \BibitemOpen
  \bibfield  {author} {\bibinfo {author} {\bibfnamefont {Greg}\ \bibnamefont
  {Ridgeway}}\ and\ \bibinfo {author} {\bibfnamefont {David}\ \bibnamefont
  {Madigan}},\ }\bibfield  {title} {\enquote {\bibinfo {title} {A sequential
  monte carlo method for bayesian analysis of massive datasets},}\ }\href@noop
  {} {\bibfield  {journal} {\bibinfo  {journal} {Data Mining and Knowledge
  Discovery}\ }\textbf {\bibinfo {volume} {7}},\ \bibinfo {pages} {301--319}
  (\bibinfo {year} {2003})}\BibitemShut {NoStop}%
\bibitem [{\citenamefont {Branicki}\ and\ \citenamefont
  {Majda}(2014)}]{branicki2014quantifying}%
  \BibitemOpen
  \bibfield  {author} {\bibinfo {author} {\bibfnamefont {M}~\bibnamefont
  {Branicki}}\ and\ \bibinfo {author} {\bibfnamefont {AJ}~\bibnamefont
  {Majda}},\ }\bibfield  {title} {\enquote {\bibinfo {title} {Quantifying
  bayesian filter performance for turbulent dynamical systems through
  information theory},}\ }\href@noop {} {\bibfield  {journal} {\bibinfo
  {journal} {Commun. Math. Sci}\ }\textbf {\bibinfo {volume} {12}},\ \bibinfo
  {pages} {901--978} (\bibinfo {year} {2014})}\BibitemShut {NoStop}%
\bibitem [{\citenamefont {Sullivan}\ \emph {et~al.}()\citenamefont {Sullivan},
  \citenamefont {Cockayne}, \citenamefont {Oates},\ and\ \citenamefont
  {Girolami}}]{sullivanbayesian}%
  \BibitemOpen
  \bibfield  {author} {\bibinfo {author} {\bibfnamefont {Tim}\ \bibnamefont
  {Sullivan}}, \bibinfo {author} {\bibfnamefont {Jon}\ \bibnamefont
  {Cockayne}}, \bibinfo {author} {\bibfnamefont {Chris}\ \bibnamefont {Oates}},
  \ and\ \bibinfo {author} {\bibfnamefont {Mark}\ \bibnamefont {Girolami}},\
  }\bibfield  {title} {\enquote {\bibinfo {title} {Bayesian probabilistic
  numerical methods},}\ }\href@noop {} {\ }\BibitemShut {NoStop}%
\bibitem [{\citenamefont {Kersting}\ and\ \citenamefont
  {Hennig}(2016)}]{kersting2016active}%
  \BibitemOpen
  \bibfield  {author} {\bibinfo {author} {\bibfnamefont {Hans}\ \bibnamefont
  {Kersting}}\ and\ \bibinfo {author} {\bibfnamefont {Philipp}\ \bibnamefont
  {Hennig}},\ }\bibfield  {title} {\enquote {\bibinfo {title} {Active
  uncertainty calibration in bayesian ode solvers},}\ }\href@noop {} {\bibfield
   {journal} {\bibinfo  {journal} {arXiv preprint arXiv:1605.03364}\ }
  (\bibinfo {year} {2016})}\BibitemShut {NoStop}%
\bibitem [{\citenamefont {De~Freitas}\ \emph {et~al.}(2001)\citenamefont
  {De~Freitas}, \citenamefont {Andrieu}, \citenamefont {H{\o}jen-S{\o}rensen},
  \citenamefont {Niranjan},\ and\ \citenamefont {Gee}}]{de2001sequential}%
  \BibitemOpen
  \bibfield  {author} {\bibinfo {author} {\bibfnamefont {Nando}\ \bibnamefont
  {De~Freitas}}, \bibinfo {author} {\bibfnamefont {C}~\bibnamefont {Andrieu}},
  \bibinfo {author} {\bibfnamefont {Pedro}\ \bibnamefont
  {H{\o}jen-S{\o}rensen}}, \bibinfo {author} {\bibfnamefont {M}~\bibnamefont
  {Niranjan}}, \ and\ \bibinfo {author} {\bibfnamefont {A}~\bibnamefont
  {Gee}},\ }\bibfield  {title} {\enquote {\bibinfo {title} {Sequential monte
  carlo methods for neural networks},}\ }in\ \href@noop {} {\emph {\bibinfo
  {booktitle} {Sequential Monte Carlo methods in practice}}}\ (\bibinfo
  {publisher} {Springer},\ \bibinfo {year} {2001})\ pp.\ \bibinfo {pages}
  {359--379}\BibitemShut {NoStop}%
\bibitem [{\citenamefont {Raissi}\ \emph
  {et~al.}(2017{\natexlab{a}})\citenamefont {Raissi}, \citenamefont
  {Perdikaris},\ and\ \citenamefont {Karniadakis}}]{raissi2017inferring}%
  \BibitemOpen
  \bibfield  {author} {\bibinfo {author} {\bibfnamefont {Maziar}\ \bibnamefont
  {Raissi}}, \bibinfo {author} {\bibfnamefont {Paris}\ \bibnamefont
  {Perdikaris}}, \ and\ \bibinfo {author} {\bibfnamefont {George~Em}\
  \bibnamefont {Karniadakis}},\ }\bibfield  {title} {\enquote {\bibinfo {title}
  {Inferring solutions of differential equations using noisy multi-fidelity
  data},}\ }\href@noop {} {\bibfield  {journal} {\bibinfo  {journal} {Journal
  of Computational Physics}\ }\textbf {\bibinfo {volume} {335}},\ \bibinfo
  {pages} {736--746} (\bibinfo {year} {2017}{\natexlab{a}})}\BibitemShut
  {NoStop}%
\bibitem [{\citenamefont {Raissi}\ \emph
  {et~al.}(2017{\natexlab{b}})\citenamefont {Raissi}, \citenamefont
  {Perdikaris},\ and\ \citenamefont {Karniadakis}}]{raissi2017numerical}%
  \BibitemOpen
  \bibfield  {author} {\bibinfo {author} {\bibfnamefont {Maziar}\ \bibnamefont
  {Raissi}}, \bibinfo {author} {\bibfnamefont {Paris}\ \bibnamefont
  {Perdikaris}}, \ and\ \bibinfo {author} {\bibfnamefont {George~Em}\
  \bibnamefont {Karniadakis}},\ }\bibfield  {title} {\enquote {\bibinfo {title}
  {Numerical gaussian processes for time-dependent and non-linear partial
  differential equations},}\ }\href@noop {} {\bibfield  {journal} {\bibinfo
  {journal} {arXiv preprint arXiv:1703.10230}\ } (\bibinfo {year}
  {2017}{\natexlab{b}})}\BibitemShut {NoStop}%
\bibitem [{\citenamefont {Archambeau}\ \emph {et~al.}(2007)\citenamefont
  {Archambeau}, \citenamefont {Cornford}, \citenamefont {Opper},\ and\
  \citenamefont {Shawe-Taylor}}]{archambeau2007gaussian}%
  \BibitemOpen
  \bibfield  {author} {\bibinfo {author} {\bibfnamefont {Cedric}\ \bibnamefont
  {Archambeau}}, \bibinfo {author} {\bibfnamefont {Dan}\ \bibnamefont
  {Cornford}}, \bibinfo {author} {\bibfnamefont {Manfred}\ \bibnamefont
  {Opper}}, \ and\ \bibinfo {author} {\bibfnamefont {John}\ \bibnamefont
  {Shawe-Taylor}},\ }\bibfield  {title} {\enquote {\bibinfo {title} {Gaussian
  process approximations of stochastic differential equations},}\ }in\
  \href@noop {} {\emph {\bibinfo {booktitle} {Gaussian Processes in
  Practice}}}\ (\bibinfo {year} {2007})\ pp.\ \bibinfo {pages}
  {1--16}\BibitemShut {NoStop}%
\bibitem [{\citenamefont {{Leike}}\ and\ \citenamefont
  {{En{\ss}lin}}(2016{\natexlab{a}})}]{2016arXiv161009018L}%
  \BibitemOpen
  \bibfield  {author} {\bibinfo {author} {\bibfnamefont {R.~H.}\ \bibnamefont
  {{Leike}}}\ and\ \bibinfo {author} {\bibfnamefont {T.~A.}\ \bibnamefont
  {{En{\ss}lin}}},\ }\bibfield  {title} {\enquote {\bibinfo {title} {{Optimal
  Belief Approximation}},}\ }\href@noop {} {\bibfield  {journal} {\bibinfo
  {journal} {ArXiv e-prints}\ } (\bibinfo {year} {2016}{\natexlab{a}})},\
  \Eprint {http://arxiv.org/abs/1610.09018} {arXiv:1610.09018 [math.ST]}
  \BibitemShut {NoStop}%
\bibitem [{\citenamefont {Bernardo}(1979)}]{bernardo1979expected}%
  \BibitemOpen
  \bibfield  {author} {\bibinfo {author} {\bibfnamefont {Jos{\'e}~M}\
  \bibnamefont {Bernardo}},\ }\bibfield  {title} {\enquote {\bibinfo {title}
  {Expected information as expected utility},}\ }\href@noop {} {\bibfield
  {journal} {\bibinfo  {journal} {The Annals of Statistics}\ ,\ \bibinfo
  {pages} {686--690}} (\bibinfo {year} {1979})}\BibitemShut {NoStop}%
\bibitem [{\citenamefont {{Wiener}}(1949)}]{1949wiener}%
  \BibitemOpen
  \bibfield  {author} {\bibinfo {author} {\bibfnamefont {N.}~\bibnamefont
  {{Wiener}}},\ }\href@noop {} {\emph {\bibinfo {title} {{Extrapolation,
  Interpolation, and Smoothing of Stationary Time Series}}}}\ (\bibinfo
  {publisher} {New York: Wiley},\ \bibinfo {year} {1949})\BibitemShut {NoStop}%
\bibitem [{\citenamefont {{Leike}}\ and\ \citenamefont
  {{En{\ss}lin}}(2016{\natexlab{b}})}]{2016PhRvE..94e3306L}%
  \BibitemOpen
  \bibfield  {author} {\bibinfo {author} {\bibfnamefont {R.~H.}\ \bibnamefont
  {{Leike}}}\ and\ \bibinfo {author} {\bibfnamefont {T.~A.}\ \bibnamefont
  {{En{\ss}lin}}},\ }\bibfield  {title} {\enquote {\bibinfo {title} {{Operator
  calculus for information field theory}},}\ }\href {\doibase
  10.1103/PhysRevE.94.053306} {\bibfield  {journal} {\bibinfo  {journal}
  {\pre}\ }\textbf {\bibinfo {volume} {94}},\ \bibinfo {eid} {053306} (\bibinfo
  {year} {2016}{\natexlab{b}})},\ \Eprint {http://arxiv.org/abs/1605.00660}
  {arXiv:1605.00660 [stat.ME]} \BibitemShut {NoStop}%
\bibitem [{\citenamefont {Wiener}(1930)}]{wiener1930generalized}%
  \BibitemOpen
  \bibfield  {author} {\bibinfo {author} {\bibfnamefont {Norbert}\ \bibnamefont
  {Wiener}},\ }\bibfield  {title} {\enquote {\bibinfo {title} {Generalized
  harmonic analysis},}\ }\href@noop {} {\bibfield  {journal} {\bibinfo
  {journal} {Acta mathematica}\ }\textbf {\bibinfo {volume} {55}},\ \bibinfo
  {pages} {117--258} (\bibinfo {year} {1930})}\BibitemShut {NoStop}%
\bibitem [{\citenamefont {{Courant}}\ \emph {et~al.}(1928)\citenamefont
  {{Courant}}, \citenamefont {{Friedrichs}},\ and\ \citenamefont
  {{Lewy}}}]{Courant1928}%
  \BibitemOpen
  \bibfield  {author} {\bibinfo {author} {\bibfnamefont {R.}~\bibnamefont
  {{Courant}}}, \bibinfo {author} {\bibfnamefont {K.O.}\ \bibnamefont
  {{Friedrichs}}}, \ and\ \bibinfo {author} {\bibfnamefont {H.}~\bibnamefont
  {{Lewy}}},\ }\bibfield  {title} {\enquote {\bibinfo {title} {\"uber die
  partiellen differenzengleichungen der mathematischen physik},}\ }\href@noop
  {} {\bibfield  {journal} {\bibinfo  {journal} {Math. Annalen}\ }\textbf
  {\bibinfo {volume} {100}},\ \bibinfo {pages} {32--74} (\bibinfo {year}
  {1928})},\ \bibinfo {note} {(English translation, with commentaries by Lax,
  P.B., Widlund, O.B., Parter, S.V., in IBM J. Res. Develop. 11
  (1967)).}\BibitemShut {Stop}%
\bibitem [{\citenamefont {{Galerkin}}(1915)}]{Galerkin1915}%
  \BibitemOpen
  \bibfield  {author} {\bibinfo {author} {\bibfnamefont {B.G.}\ \bibnamefont
  {{Galerkin}}},\ }\bibfield  {title} {\enquote {\bibinfo {title} {On
  electrical circuits for the approximate solution of the laplace equation},}\
  }\href@noop {} {\bibfield  {journal} {\bibinfo  {journal} {Vestnik Inzh.}\
  }\textbf {\bibinfo {volume} {19}},\ \bibinfo {pages} {897--908} (\bibinfo
  {year} {1915})}\BibitemShut {NoStop}%
\end{thebibliography}%

\end{document}